\documentclass[prd,showpacs]{revtex4}
\begin{document}
\title{
Effective electromagnetic lagrangian at finite temperature and density in the electroweak model
      }
\author{Andrea Erdas}
\affiliation{
 Department of Physics, Loyola University Maryland, 4501 North Charles Street
       Baltimore, Maryland 21210, USA}
\email{aerdas@loyola.edu}
\date{June 24, 2010}
\begin {abstract} 
Using the exact propagators
in a constant magnetic field, the effective electromagnetic lagrangian at finite temperature and density is calculated
to all orders in the field strength $B$ within the framework of the complete electroweak model, in the weak coupling limit. 
The partition function and free energy are obtained explicitly and the finite temperature effective coupling is derived 
in closed form. Some implications of this result, potentially interesting to astrophysics and cosmology, are discussed. 
\end {abstract}
\pacs{11.10.Wx, 05.30.Fk, 05.30.Jp, 13.40.-f}
\maketitle
%%%%%%%%%%%%%%%%%%%%%%%%%%%%%%%%%%%%%%%%%%%%%%%%%%%%%%%%%%%%%%%%%%%%%
\section{Introduction}
%%%%%%%%%%%%%%%%%%%%%%%%%%%%%%%%%%%%%%%%%%%%%%%%%%%%%%%%%%%%%%%%%%%%%
Large magnetic fields are present in a variety of 
astrophysical sites like supernovae, neutron stars and white dwarfs, and even larger fields 
can arise in supernovae explosions or coalescing neutron stars. The remnants of such astrophysical cataclysms
are called magnetars, young neutron stars with magnetic fields $10^{14}-10^{16}$ G \cite{Duncan,Thompson1,Thompson2}.
It has been argued that during the electroweak
phase transition local magnetic fields much stronger than those of a magnetar could have existed, with field strength
as high as
$10^{22}-10^{24}$ G \cite{Brandenburg,Joyce,grasso}. In many of these situations the thermal and density effects 
of the medium are very important and must be considered, and many of the thermodynamical quantities that characterize
the medium, like free energy, thermodynamical partition function and effective potential are closely related to the 
effective action and effective lagrangian.

The work of Heisenberg and Euler \cite{heisenberg}, who calculated the one-loop vacuum 
effective lagrangian for spinor QED in a uniform electromagnetic background field, and that of Weisskopf 
\cite{Weisskopf:1996bu},
who calculated the analogous quantity for scalar QED, were published 
many years ago and are the first examples of what we now call low energy effective field theory. 
These pioneering papers lead to a number of important physical insights
and applications: light-light scattering in QED \cite{karplus}, pair production from vacuum in the presence 
of an electric field \cite{schwinger,Dunne:2008kc,Hebenstreit:2009km} 
and vacuum birefringence \cite{dittrich2}, among others. 
The one-loop QED effective lagrangian at finite temperature and density has been investigated in magnetic field 
background \cite{elmfors1,elmfors2,elmfors3,chodos},
in electric field background \cite{Kim:2008em,Kim:2010qq},
in general background fields \cite{Das:2009ci,Das:2009vs}, and 
is very relevant and closely related to many physical phenomena such as, for example, the Casimir effect. 
In the former the time component of the momentum four vector, over which we integrate, 
takes on only discrete values for a fixed temperature, 
while in the latter an analogous substitution takes place in a space component of the momentum vector for
a fixed distance between the plates. Recently there has been much interest \cite{Teo:2008ah,Lim:2009zza,Milton:2010yw}
in studying the finite temperature Casimir effect in higher dimensional space-time models with compactified extra dimensions,
like the Randall-Sundrum models \cite{Frank:2007jb,Rypestol:2009pe}.

A literature search reveals that the effective lagrangian in a background magnetic field at finite temperature and density
has been studied within the framework of QED \cite{schwinger,Dunne:2008kc,Hebenstreit:2009km,Kim:2008em}, or within the 
framework of the electroweak model but prior to the breaking of the electroweak symmetry, when the magnetic fields that are
present belong to the $U(1)$ group of hypercharge and hence are called hypermagnetic fields
\cite{Cannellos:2002ex,Piccinelli:2004eu,Sanchez:2006tt}.
The finite temperature and density QED effective lagrangian for the complete electroweak model after symmetry breaking,
however, has never been obtained. When magnetic fields are much larger than $B_e=m^2_e/e\simeq 4.414 \times 10^{13}$ G,
 where $m_e$ and $-e$ are the 
electron mass and charge respectively, the full electroweak model must be used, 
since electroweak magnetism \cite{Ambjorn:1992ca} becomes important.
In this paper, using Schwinger's proper time method \cite{schwinger}, I calculate the effective electromagnetic
lagrangian of the complete electroweak model for a thermal environment treated exactly in 
the external constant magnetic field and in the weak coupling limit, i.e. with no virtual photons present.

In Section \ref{2} the notation for the fermion, gauge boson and scalar thermal propagators in a constant magnetic field
background is presented. 
In Section \ref{3} the finite temperature and density effective electromagnetic lagrangian
for the complete electroweak model is obtained.
An extended discussion of several implications of my results is presented in Section \ref{4}. In the
Appendix I evaluate the next to leading order correction to the
Heisenberg-Euler effective lagrangian and to the Weisskopf effective lagrangian 
of scalar QED in the case of strong magnetic field.
%%%%%%%%%%%%%%%%%%%%%%%%%%%%%%%%%%%%%%%%%%%%%%%%%%%%%%%%%%%%%%%%%%%%%
\section{ Vacuum and thermal propagators in a constant magnetic field}
\label{2}
The metric used in this paper is
$g^{\mu \nu} = \mathrm{diag}(+1,-1,-1,-1)$ and the $z$-axis points in the direction of the constant 
magnetic field $\mathbf{B}$. Therefore the electromagnetic
field strength tensor $F^{\mu \nu}$ has only two nonvanishing components $F^{21}= -F^{12} = B$. 

I start by recalling the expressions for the charged lepton
$S_0(x',x'')$ \cite{schwinger,dittrich}, $W$ boson
$G^{\mu \nu}_0(x',x'')$ and scalar vacuum propagators $D_0(x',x'')$  \cite{erdasfeld} in a constant
magnetic field. These propagators are written in the Feynman gauge and derived using Schwinger's proper time method:
\begin{equation}
S_0(x',x'')=\Omega(x',x'')\!\int{d^4k\over (2\pi)^4}e^{-ik\cdot (x'-x'')}
S_0(k) \quad , 
\label{s0x}
\end{equation}
\begin{equation}
G^{\mu\nu}_0(x',x'')=\Omega(x',x'')\!\int{d^4k\over (2\pi)^4}e^{-ik\cdot
(x'-x'')}
G^{\mu\nu}_0(k) \quad ,
\label{g0x}
\end{equation}
\begin{equation}
D_0(x',x'')=\Omega(x',x'')\!\int{d^4k\over (2\pi)^4}e^{-ik\cdot (x'-x'')}
D_0(k) \quad .
\label{d0x}
\end{equation}
The translationally 
invariant parts of the propagators are
\begin{equation}
S_0(k) =\int_0^\infty \!\!{ds\over\cos eBs} 
{\exp{\left[-is\left(m^2-k^2_{\parallel}-
k^2_{\perp}{\tan eBs\over eBs}-i\epsilon\right)\right]}}
\left[(m+
\not\! k_{\parallel})e^{-ieBs\sigma_3}+
{\not\! k_{\perp}\over \cos eBs}
\right] , 
\label{S0}
\end{equation}
\begin{equation}
G^{\mu \nu}_0(k)=
-\int_0^\infty \!\!{ds\over\cos eBs}\,
{\exp{\left[-is\left(M^2-k^2_{\parallel}-
k^2_{\perp}{\tan eBs\over eBs}-i\epsilon\right)\right]}}
[g^{\mu \nu}_{\parallel}
-(e^{2eFs})^{\mu\nu}],
\label{G0}
\end{equation}
\begin{equation}
D_0(k)=
\int_0^\infty \!\!{ds\over\cos eBs}\,
{\exp{\left[-is\left(M^2-k^2_{\parallel}-
k^2_{\perp}{\tan eBs\over eBs}-i\epsilon\right)\right]}},
\label{D0}
\end{equation}
where $-e$ and
$m$ are the charge and mass of the charged lepton,
$M$ is the $W$-mass, and the $-i\epsilon$ prescription for the propagators is essential for the
convergence of the $s$ integrals. 
It is convenient to use the notation
\begin{equation}
a^\mu_{\parallel}=(a^0,0,0,a^3), \quad
a^\mu_{\perp}=(0,a^1,a^2,0)
\label{amu}
\end{equation}
and 
\begin{equation}(a b)_\parallel=a^0\,b^0-a^3\,b^3, \quad
(a b)_\perp=-a^1\,b^1-a^2\,b^2
\label{adotb}
\end{equation}
for arbitrary four-vectors $a$ and $b$. Using this notation I write the metric tensor as
\begin{equation}
g^{\mu\nu}=g^{\mu\nu}_\parallel+g^{\mu\nu}_\perp\quad .
\label{g}
\end{equation} 
The $4\times 4$ matrix $\sigma_3$ that appears in the charged lepton propagator (\ref{S0}), 
is given by
\begin{equation} 
\sigma_3=
{i \over 2}[\gamma^1, \gamma^2] \quad .
\label{sigma3}
\end{equation} 
When writing the $W$ propagator (\ref{G0}), I use the notation
\begin{equation}
\left(e^{2eFs}\right)^{{\mu}\nu}=
-g^{\mu\nu}_{\perp} \cos{(2eBs)}
+{F^{\mu\nu}\over B}  \sin{(2eBs)} \quad .
\label{efmunu}\end{equation}
I choose the electromagnetic vector potential to be
$A_\mu=-{1\over2}F_{\mu \nu}x^\nu$ and therefore
the phase factor which appears in Equations (\ref{s0x}-\ref{d0x}) is given by
\cite{dittrich}
\begin{equation}
\Omega(x',x'')=\exp\left(
-i {e\over 2}x'_\mu F^{\mu \nu} x''_\nu
\right)  \quad .
\label{omega}
\end{equation}
Notice that, as $B\rightarrow 0$, the phase factor $\Omega \rightarrow 1$ and the translationally invariant part 
of the propagators take the standard form found in many quantum field theory textbooks
\begin{equation}
S_0(k)\rightarrow i{{\not\!\, k}+m\over k^2-m^2+i\epsilon}  \quad , \quad G^{\mu \nu}_0(k)
\rightarrow {-ig^{\mu\nu}\over k^2-M^2+i\epsilon} \quad , \quad D_0(k)
\rightarrow {i\over k^2-M^2+i\epsilon} \quad .
\label{vac_props}
\end{equation}

At finite temperature and density we need to replace the vacuum propagators in Equations (\ref{S0} - \ref{D0}) 
with thermal propagators. The real-time thermal propagators $S(x',x'')$, 
$G^{\mu \nu}(x',x'')$ and $D(x',x'')$ are easily constructed starting from the 
proper-time form of the vacuum propagators
\begin{equation}
S(x',x'')=\Omega(x',x'')\!\int{d^4k\over (2\pi)^4}e^{-ik\cdot (x'-x'')}
S(k) \quad , 
\label{sx}
\end{equation}
\begin{equation}
G^{\mu\nu}(x',x'')=\Omega(x',x'')\!\int{d^4k\over (2\pi)^4}e^{-ik\cdot
(x'-x'')}
G^{\mu\nu}(k) \quad ,
\label{gx}
\end{equation}
\begin{equation}
D(x',x'')=\Omega(x',x'')\!\int{d^4k\over (2\pi)^4}e^{-ik\cdot (x'-x'')}
D(k) \quad ,
\label{dx}
\end{equation}
where $S(k)$, $G^{\mu \nu}(k)$ and $D(k)$ are defined in terms of the translationally 
invariant parts of the vacuum propagators, $S_0(k)$, $G^{\mu \nu}_0(k)$ and $D_0(k)$, and of the fermion and
boson occupation numbers $f_F(k^0)$  and $f_B(k^0)$
\begin{equation}
S(k)=S_0(k)-f_F(k^0)\Bigl[S_0(k)+S_0^\ast (k)\Bigr]\quad ,
\label{sk}
\end{equation}
\begin{equation}
G^{\mu\nu}(k)=G^{\mu\nu}_0(k)+f_B(k^0)\Bigl[G^{\mu\nu}_0(k)+
{G^{\mu\nu}_0}^\ast(k)\Bigr]\quad ,
\label{gk}
\end{equation}
\begin{equation}
\Delta(k)=\Delta_0(k)+f_B(k^0)\Bigl[\Delta_0(k)+\Delta_0^\ast (k)\Bigr]\quad .
\label{dk}
\end{equation}
Notice that in Equations (\ref{sk}-\ref{dk}) the pieces proportional
to the occupation numbers represent the thermal parts of the propagators. 
The fermion occupation number at temperature $T$ and chemical potential $\mu$ is defined as 
\begin{equation}
f_F(k^0)=f_F^+(k^0)\theta(k^0)+f_F^-(k^0)\theta(-k^0)
\label{fF}
\end{equation}
with
\begin{equation}
f_F^\pm(k^0)={1\over e^{\pm\beta(k^0-\mu)}+1}\quad ,
\end{equation}
and the boson occupation number $f_B(k^0)$ is defined as
\begin{equation}
f_B(k^0)={1\over e^{\beta|k^0|}-1}\quad ,
\end{equation}
with $\beta = T^{-1}$. These thermal propagators cannot be used naively for perturbative expansions.
The reason is that there is a delta function hidden in the thermal part of the propagators, and the overlap of
delta functions with coinciding arguments on several internal legs leads to expressions that are not well
defined. This problem is solved in the real-time formalism of finite temperature field theory by doubling the 
field degrees of freedom and introducing thermal "ghost" fields. Of course only the "physical" fields occur
on external lines of the Green's functions. Each field and its thermal "ghost" are therefore grouped into 
thermal doublets, and the thermal propagators become $2\times 2$ matrices with off-diagonal elements.
I will indicate the matrix propagators with a tilde, and below I write the matrix form of the 
translationally invariant part of the fermion thermal propagator
\begin{equation}
{\tilde S}(k)=U_F(k^0)
\left(
\begin{array}{cc}
S_0(k) & 0     \\
0      & -S_0^\ast (k) 
\end{array}
\right)
U_F^T(k^0)\quad ,
\label{stildek}
\end{equation}
where
\begin{equation}
U_F(k^0)=
\left(
\begin{array}{cc}
\cos\vartheta_F (k^0) & -\sin\vartheta_F (k^0)    \\
\sin\vartheta_F (k^0) & \cos\vartheta_F (k^0)
\end{array}
\right)\quad ,
\label{UF}
\end{equation}
and
\begin{equation}
\sin\vartheta_F (k^0) =\theta(k^0)\sqrt{f_F^+(k^0)}-\theta(-k^0)\sqrt{f_F^-(-k^0)}\quad ,
\end{equation}
\begin{equation}
\cos\vartheta_F (k^0) =\theta(k^0)\sqrt{1-f_F^+(k^0)}-\theta(-k^0)\sqrt{1-f_F^-(-k^0)}\quad .
\end{equation}
Here, $U_F^T(k^0)$ is the transpose of $U_F(k^0)$. The ${\tilde S}(k)_{11}$ component is, of course,
the same as the propagator in Equation (\ref{sk}) and the other components are only used 
in higher loop calculations. The matrix forms of the translationally invariant part of the charged vector and 
scalar field thermal propagators are
\begin{equation}
{\tilde G}^{\mu\nu}(k)=U_B(k^0)
\left(
\begin{array}{cc}
G^{\mu\nu}_0(k) & 0     \\
0      & {G^{\mu\nu}_0}^\ast (k) 
\end{array}
\right)
U_B(k^0)\quad ,
\label{gtildek}
\end{equation}
\begin{equation}
{\tilde D}(k)=U_B(k^0)
\left(
\begin{array}{cc}
D_0(k) & 0     \\
0      & D_0^\ast (k) 
\end{array}
\right)
U_B(k^0)\quad ,
\label{dtildek}
\end{equation}
with
\begin{equation}
U_B(k^0)=
\left(
\begin{array}{cc}
\sqrt{1+f_B(k^0)} & \sqrt{f_B(k^0)}   \\
\sqrt{f_B(k^0)}& \sqrt{1+f_B(k^0)}
\end{array}
\right)\quad .
\label{UB}
\end{equation}
The ${\tilde G}^{\mu\nu}(k)_{11}$ and ${\tilde D}(k)_{11}$ components are
the same as the propagators in Equations (\ref{gk}) and (\ref{dk}) and the other components are only needed 
in higher loop calculations.
%%%%%%%%%%%%%%%%%%%%%%%%%%%%%%%%%%%%%%%%%%%%%%%%%%%%%%%%%%%%%%%%%%%%%
\section{ Finite temperature and density effective electromagnetic lagrangian}
\label{3}
In vacuum the effective EM action $S_{eff}$ of the electroweak model at the one-loop level can be written as
\begin{equation}
S_{eff}=\int d^4x \,\,{\cal{L}}_{eff},
\label{action}
\end{equation}
and the effective EM lagrangian ${\cal{L}}_{eff}$ is given by
\begin{equation}
{\cal{L}}_{eff}={\cal{L}}^{(0)} + {\cal{L}}^{(1)},
\label{eff_lag}
\end{equation}
where the tree level part in the case of a constant magnetic field is the classical lagrangian density
\begin{equation}
{\cal{L}}^{(0)} = -{1\over 2} B^2,
\label{L0}
\end{equation}
and ${\cal{L}}^{(1)}$ is the one-loop quantum correction.
In the Feynman gauge, all the charged fields of the electroweak model contribute to the one-loop effective lagrangian
\begin{equation}
{\cal{L}}^{(1)}=\sum_f {\cal{L}}_f + {\cal{L}}_W + \sum_s{\cal{L}}_s\,,
\label{eff_lag2}
\end{equation}
where ${\cal{L}}_f$, ${\cal{L}}_s$, ${\cal{L}}_W$ indicate the contributions of fermion, 
scalar and $W$-fields respectively, and we need to sum over the quarks and charged leptons and over the 
non-physical charged scalars. In a medium at finite temperature and density, 
we still use Equation (\ref{eff_lag2}) to calculate ${\cal{L}}^{(1)}$,
but the contributions of the various fields are evaluated using thermal propagators instead of vacuum propagators,
and we find that ${\cal{L}}^{(1)}$ is the sum of a vacuum part ${\cal{L}}^0$ and a thermal part 
${\cal{L}}^T$ 
\begin{equation}
{\cal{L}}^{(1)} = {\cal{L}}^0+ {\cal{L}}^T\,.
\label{Lf2}
\end{equation}
The same is true for ${\cal{L}}_f={\cal{L}}^0_f+ {\cal{L}}^T_f$, for ${\cal{L}}_W$ and for ${\cal{L}}_s$ and,
clearly, ${\cal L}^{0}=\sum_f{\cal{L}}_f^0+ {\cal{L}}_W^0 +\sum_s{\cal{L}}_s^0 $ and 
${\cal L}^{T}=\sum_f{\cal{L}}_f^T+ {\cal{L}}_W^T +\sum_s{\cal{L}}_s^T$.

${\cal{L}}_f$ was 
calculated in Refs. \cite{elmfors1,elmfors2} using the solutions to the Dirac equation in a uniform
magnetic field to construct the fermion thermal propagator. 
In this paper I calculate ${\cal{L}}_f$ using the thermal propagator of Equation (\ref{sx}), which is constructed using the 
Schwinger proper time method, and use the following identity \cite{elmfors1,elmfors2} to evaluate 
the contribution to the effective lagrangian of a charged lepton field whose thermal propagator is
$S(x',x'')$
\begin{equation}
{\partial{\cal{L}}_f\over \partial m} ={\rm Tr}\,S(x,x)\,,
\label{Lf}
\end{equation}
where the trace is over spinor indices. The generalization to quarks is straightforward. 
After taking the spinor trace, I find
\begin{equation}
{\partial{\cal{L}}^0_f\over \partial m} =\,4m\int{d^4k\over (2\pi)^4}
\int_0^\infty \!\!{ds}\,\,
{\cal E}(m)
\label{Lf0}
\end{equation}
and
\begin{equation}
{\partial{\cal{L}}_f^T\over \partial m} =\,-4m\int{d^4k\over (2\pi)^4}f_F(k^0)
\int_0^\infty \!\!{ds}\,\,
\left[{\cal E}(m)+{\cal E}^\ast(m)\right]\,,
\label{LfT}
\end{equation}
where I introduce the notation
\begin{equation}
{\cal E}(m)={\exp{\left[-is\left(m^2-k^2_{\parallel}-
k^2_{\perp}{\tan eBs\over eBs}-i\epsilon\right)\right]}}.
\label{E}
\end{equation}
After the straightforward integration over the four variables $k^\mu$ and a rotation of the $s$ contour
to the negative imaginary axis, I obtain immediately the derivative of the vacuum part
\begin{equation}
{\partial{\cal{L}}_f^0\over \partial m} ={m\over 4\pi^2}
\int_0^\infty \!{ds\over s^2}\,\,
e^{-s m^2}\, eBs\coth (eBs)\,,
\label{Lf02}
\end{equation}
and find
\begin{equation}
{\cal{L}}_f^0 =-{1\over 8\pi^2}
\int_0^\infty \!{ds\over s^3}\,\,
e^{-s m^2}\left[eBs \coth (eBs)-1-{(eBs)^2\over 3}\right]\,,
\label{Lf03}
\end{equation}
which has been renormalized by adding a second order polynomial in $eB$ and reproduces 
the old result by Schwinger \cite{schwinger}. This charge and wave function renormalization procedure
leaves $eB$ invariant and produces a contribution of ${2\alpha\over 3\pi} $to the QED $\beta$ function from 
our fermion field.

In order to obtain the thermal part ${\cal{L}}_f^T$, I start by integrating over the three $k^i$ variables 
to find
\begin{equation}
{\partial{\cal{L}}_f^T\over \partial m} =-{m \over 2\pi^{5/2}}
\int_{-\infty}^{+\infty} dk^0 f_F(k^0)\,{\rm Re}\int_0^\infty \!ds \,{\cal I}_f(s)\,,
\label{LfT2}
\end{equation}
where
\begin{equation}
{\cal I}_f(s) =
{e^{-is (m^2-k_0^2-i\epsilon)}\over (is)^{3/2}}\,\,
\, eBs \cot (eBs)\,.
\label{Is}
\end{equation}
Next I need to rotate the $s$ integration contour to the positive imaginary axis, and must be careful with its convergence 
and analytic structure. After the contour rotation, I find only contribution from $|k^0| > m$,
as expected, and, since the rotated contour crossed the poles of ${\cal I}(s)$, I need to include a sum 
over the residues of ${\cal I}_f(s)$ at its poles $s={n\pi\over eB}$ 
(for $n\ge 1$)
\begin{equation}
2\pi i \,{\rm Res}\left[{\cal I}_f(s={n\pi\over eB})\right] = 2\left({\pi eB\over n}\right)^{1/2}
\exp\left[{i n\pi \over eB}(k_0^2-m^2)-{i \pi \over 4}\right].
\label{res}
\end{equation}
I obtain the following $s$-integral and sum over the real parts of the residues of ${\cal I}(s)$
\begin{eqnarray}
{\partial{\cal{L}}_f^T\over \partial m} &=&{m \over 2\pi^{5/2}}
\int_{-\infty}^{+\infty} dk^0 f_F(k^0)\theta(k_0^2-m^2)\left[\int_0^\infty \!{ds\over s^{3/2}}
e^{-s (k_0^2-m^2)}eBs \coth (eBs)\right.
\nonumber \\
&-&
 \left. 2\sum^\infty_{n=1}\left({\pi eB\over n}\right)^{1/2}
\cos\left({\pi\over 4}-{n\pi\over eB}(k_0^2-m^2)\right)\right]\,,
\label{LfT3}
\end{eqnarray}
and find the fermion contribution to the effective Lagrangian
\begin{eqnarray}
{\cal{L}}_f^T &=&{1\over 4\pi^{5/2}}
\int_{-\infty}^{+\infty} dk^0 f_F(k^0)\theta(k_0^2-m^2)\left[\int_0^\infty \!{ds\over s^{5/2}}
e^{-s (k_0^2-m^2)}eBs \coth (eBs)\right.
\nonumber \\
&-&
 \left. {2\over \pi^{1/2}}\sum^\infty_{n=1}\left({eB\over n}\right)^{3/2}
\sin\left({\pi\over 4}-{n\pi\over eB}(k_0^2-m^2)\right)\right]\,,
\label{LfT4}
\end{eqnarray}
which agrees with the result obtained by a different method in Refs. \cite{elmfors1,elmfors2}.
Using the form of ${\cal{L}}_f^T$ given in Equation (\ref{LfT4}), it is shown in Ref. \cite{elmfors1} that a 
fermion-antifermion plasma in a magnetic field exhibits de Haas-van Alphen oscillations in the limit
where $T=0$, and $eB\ll \mu^2 - m^2 \ll m^2$. The second term in Equation (\ref{LfT4}) is responsible for the 
oscillatory behavior, and the oscillation frequency 
agrees with the one derived by Onsager for the de Haas-van Alphen effect.
We can write ${\cal{L}}_f^T$ in another form that seems less transparent of its physical content
but clearly displays its connection to the partition
function $Z_f(B,T,\mu)$ of a relativistic fermion-antifermion gas in a magnetic field in a sufficiently large 
quantization volume $V$.
Starting from Equation (\ref{LfT}), I integrate over the $k_{\perp}$ variables and do a Wick rotation of 
the $s$ variable to obtain
\begin{equation}
{\partial{\cal{L}}_f^T\over \partial m} ={m \over 2\pi^{3}}
\int d^2k_{\parallel} f_F(k^0)\,{\rm Re}\int_0^\infty \!ds \,\exp\left[-s(m^2-k^2_{\parallel} -i\epsilon)\right]
\,ieB\coth(eBs)\,.
\label{LfT5}
\end{equation}
Next I use the following identity
\begin{equation}
\coth(eBs)=\sum_{n=0}^\infty\sum_{\lambda = 1}^2e^{-2eBs(n+\lambda - 1)}\,,
\label{cot}
\end{equation}
and do the $s$ integration, then use
\begin{equation}
{\rm Re}{i\over x -i\epsilon}=-\pi\delta(x)\,,
\label{delta}
\end{equation}
and finally integrate over $k^0$ to obtain
\begin{equation}
{\partial {\cal{L}}_f^T\over \partial m} =-m{eB\over 4\pi^{2}}\sum_{n=0}^\infty\sum_{\lambda = 1}^2
\int_{-\infty}^\infty dk {f_F(\omega_{n,\lambda})+f_F(-\omega_{n,\lambda})\over \omega_{n,\lambda}}\,,
\label{LfT6}
\end{equation}
where
\begin{equation}
\omega_{n,\lambda}=\sqrt{m^2+k^2+2eB(n+\lambda-1)}
\label{omeg}
\end{equation}
are the standard Landau energy levels for a spin-${1\over 2}$ fermion and $\lambda$ is the spin quantum number.
${\cal{L}}_f^T$ is now obtained immediately with an integration by parts with respect to $k$
\begin{equation}
{\cal{L}}_f^T ={eB\over 4\pi^{2}}
\sum_{n=0}^\infty\sum_{\lambda = 1}^2
\int_{-\infty}^\infty dk {k^2\over \omega_{n,\lambda}}[f_F(\omega_{n,\lambda})+f_F(-\omega_{n,\lambda})]\,.
\label{LfT7}
\end{equation}
With ${\cal{L}}_f^T$ written in this form, its connection
to the partition function of a relativistic fermion-antifermion plasma placed
in a box of volume $V$ and in the presence of an external magnetic field $B$
is apparent \cite{elmfors1,elmfors2,elmfors3}
\begin{equation}
{\cal{L}}_f^T ={\ln Z_f(B,T,\mu)\over \beta V}\,,
\label{Z}
\end{equation}
and so is ${\cal{L}}_f^T$ connection to the free energy $F_f(B,T,\mu)$ of the fermion-antifermion plasma
\begin{equation}
{\cal{L}}_f^T ={ F_f(B,T,\mu)\over V}\,.
\label{F}
\end{equation}
The strong field limit is more easily obtained using Equation (\ref{LfT7}). For $eB\gg T^2,m^2,\mu^2$ only the lowest 
Landau level contributes to ${\cal{L}}_f^T$, and we find
\begin{equation}
{\cal{L}}_f^T ={eB\over 4\pi^{2}}
\int_{-\infty}^\infty dk {k^2\over \omega_{0,1}}[f_F(\omega_{0,1})+f_F(-\omega_{0,1})]\,,
\label{LfhighT}
\end{equation}
where the energy of the lowest Landau level is $\omega_{0,1}=\sqrt {k^2+m^2}$. It is shown in Ref. \cite{elmfors2},
through a rather lengthy calculation, that also Equation (\ref{LfT4}) yields the same result.
A complete analysis of ${\cal{L}}_f^T$ for various large or small limits of the dimensionful parameters $T,B,m$ and $\mu$,
will be done in Sec. \ref{4}.

The contribution of the $W$ fields to the effective lagrangian is obtained using the following
\begin{equation}
{\partial{\cal{L}}_W\over \partial M^2} =\,G^\mu_{\,\mu}(x,x)\,,
\label{LW}
\end{equation}
where $G^{\mu \nu}(x',x'')$ is the thermal propagator of Equation (\ref{gx}). After taking the trace, the derivatives of the 
vacuum and thermal part of ${\cal{L}}_W$ are found to be
\begin{equation}
{\partial{\cal{L}}^0_W\over \partial M^2} =-\int{d^4k\over (2\pi)^4}
\int_0^\infty \!\!{ds}\,\,
{\cal E}(M){2+2\cos (2eBs)\over \cos (eBs)},
\label{LW0}
\end{equation}
and \begin{equation}
{\partial{\cal{L}}^T_W\over \partial M^2} =-\int{d^4k\over (2\pi)^4}f_B(k^0)
\int_0^\infty \!\!{ds}\,\,
\left[{\cal E}(M)+{\cal E}^\ast(M)\right]{2+2\cos (2eBs)\over \cos (eBs)}\,,
\label{LWT}
\end{equation}
where $\cal E$ is defined in Equation (\ref{E}). I do the $k^\mu$ integration and a Wick rotation of the
$s$ variable to obtain
\begin{equation}
{\partial{\cal{L}}^0_W\over \partial M^2} =-{1\over (2\pi)^2}
\int_0^\infty \!\!{ds\over s^2}\,\,
e^{-sM^2}{eBs \coth (eBs) \cosh (eBs)},
\label{LW01}
\end{equation}
and find the vacuum contribution of the $W$ fields to the effective lagrangian \cite{erdasfeld}
\begin{equation}
{\cal{L}}^0_W ={1\over (2\pi)^2}
\int_0^\infty \!\!{ds\over s^3}\,\,
e^{-sM^2}\left[{eBs \coth (eBs) \cosh (eBs)}-1-{5\over 6}(eBs)^2\right],
\label{LW02}
\end{equation}
renormalized by adding a second order polynomial in $eB$.

I now use Equation (\ref{LWT}) and integrate over the $k^i$ variables to find
\begin{equation}
{\partial{\cal{L}}_W^T\over \partial M^2} =-{1 \over 4\pi^{5/2}}
\int_{-\infty}^{+\infty} dk^0 f_B(k^0)\,{\rm Re}\int_0^\infty \!ds \,{\cal I}_W(s)\,,
\label{LWT2}
\end{equation}
where
\begin{equation}
{\cal I}_W(s) =
{e^{-is (M^2-k_0^2-i\epsilon)}\over (is)^{3/2}} {eBs\over \sin(eBs)}[1+\cos(2eBs)]\,.
\label{IW}
\end{equation}
Next I deform the $s$ integration contour to the imaginary axis and cross the 
poles of ${\cal I}_W(s)$ at $s={n\pi\over eB}$ in the process. 
After including the contributions of the residues of these poles,
I obtain
\begin{eqnarray}
{\partial{\cal{L}}_W^T\over \partial M^2} &=&{1\over 8\pi^{5/2}}\sum_{\lambda' = -1}^{+1}(2-|\lambda'|)
\int_{-\infty}^{+\infty} dk^0 f_B(k^0)\theta[k_0^2-M^2+(2|\lambda'|-1)eB]
\left[\int_0^\infty \!{ds\over s^{3/2}}e^{-s (k_0^2-M^2)}\right.
\nonumber \\
&\times &
\left.
 e^{-2\lambda' eBs}{eBs\over \sinh (eBs)}
-2\sum^\infty_{n=1}(-1)^n\left({\pi eB\over n}\right)^{1/2}
\cos\left({\pi\over 4}-{n\pi\over eB}(k_0^2-M^2)\right)\right]\,,
\label{LWT3}
\end{eqnarray}
and find
\begin{eqnarray}
{\cal{L}}_W^T &=&{1\over 8\pi^{5/2}}\sum_{\lambda' = -1}^{+1}(2-|\lambda'|)
\int_{-\infty}^{+\infty} dk^0 f_B(k^0)\theta[k_0^2-M^2+(2|\lambda'|-1)eB]
\left[\int_0^\infty \!{ds\over s^{5/2}}e^{-s (k_0^2-M^2)}\right.
\nonumber \\
&\times &
\left.
 e^{-2\lambda' eBs}{eBs\over \sinh (eBs)}
-{2\over \pi^{1/2}}\sum^\infty_{n=1}(-1)^n\left({eB\over n}\right)^{3/2}
\sin\left({\pi\over 4}-{n\pi\over eB}(k_0^2-M^2)\right)\right]\,.
\label{LWT4}
\end{eqnarray}
The $W$ contribution to the effective electromagnetic lagrangian at finite temperature 
is one of the main results of this paper. ${\cal{L}}_W^T$ can be written in a different form,
that is obtained using steps similar to those outlined in Equations (\ref{LfT5} - \ref{LfT7})
\begin{equation}
{\cal{L}}_W^T ={eB\over 4\pi^{2}}
\sum_{n=0}^\infty\sum_{\lambda' = -1}^{+1}(2-|\lambda'|)
\int_{-\infty}^\infty dk {k^2\over E_{n,\lambda'}}[f_B(E_{n,\lambda'})+f_B(-E_{n,\lambda'})]\,,
\label{LfW5}
\end{equation}
where the spin quantum number $\lambda'$ takes the three values $0,\pm 1$ and the Landau 
energy levels for a spin-$1$ boson are
\begin{equation}
E_{n,\lambda'} =\sqrt{M^2+k^2+(2n+2\lambda' + 1)eB}\,.
\label{Enl}
\end{equation}

Last I obtain the scalar field contribution to the effective lagrangian using the following
\begin{equation}
{\partial{\cal{L}}_s\over \partial M^2} =-D(x,x)\,,
\label{Ls}
\end{equation}
where $D(x',x")$ is the scalar propagator. The vacuum part of ${\cal{L}}_s$ is \cite{Weisskopf:1996bu,erdasfeld}
\begin{equation}
{\cal{L}}^0_s ={1\over (4\pi)^2}
\int_0^\infty \!\!{ds\over s^3}\,\,
e^{-sM^2}\left[{eBs \over\sinh (eBs)}-1+{1\over 6}(eBs)^2\right],
\label{Ls01}
\end{equation}
renormalized by adding a second order polynomial in $eB$.
The thermal part of the scalar field contribution is \cite{elmfors2}
\begin{eqnarray}
{\cal{L}}_s^T &=&{1\over 8\pi^{5/2}}
\int_{-\infty}^{+\infty} dk^0 f_B(k^0)\theta(k_0^2-M^2-eB)
\left[\int_0^\infty \!{ds\over s^{5/2}}e^{-s (k_0^2-M^2)}\right.
\nonumber \\
&\times &
\left.
{eBs\over \sinh (eBs)}
-{2\over \pi^{1/2}}\sum^\infty_{n=1}(-1)^n\left({eB\over n}\right)^{3/2}
\sin\left({\pi\over 4}-{n\pi\over eB}(k_0^2-M^2)\right)\right]\,,
\label{LsT1}
\end{eqnarray}
and I can write it also as
\begin{equation}
{\cal{L}}_s^T ={eB\over 4\pi^{2}}
\sum_{n=0}^\infty
\int_{-\infty}^\infty dk {k^2\over E_{n,0}}[f_B(E_{n,0})+f_B(-E_{n,0})]\,,
\label{LsT2}
\end{equation}
where $E_{n,0}$ are the Landau energy levels of Equation (\ref{Enl}) with $\lambda' = 0$ for the spin-$0$ scalar 
field. Equation (\ref{LsT2}) shows the connection of ${\cal{L}}_s^T$ to the partition function $Z_s(B,T)$ and free energy
$F_s(B,T)$ of a spin-0 boson plasma placed in a volume $V$ and in the presence of a constant magnetic field $B$
\begin{equation}
{\cal{L}}_s^T ={\ln Z_s(B,T)\over \beta V}={ F_s(B,T)\over V}\,.
\label{Zs}
\end{equation}

In the Feynman gauge the quantity $\sum{\cal{L}}_s$, which is the contribution of the non-physical charged scalars 
to the finite temperature effective lagrangian ${\cal L}^{(1)}$ at one loop,
is obtained by adding the vacuum and thermal contributions coming from the Goldstone (scalar boson) field of mass $M$,
and the two ghost fields (scalar fermions of mass $M$). The contribution
of each of the ghost fields is opposite to that of the Goldstone boson, and therefore I find
\begin{eqnarray}
{\cal{L}}_W^T +\sum_s{\cal{L}}_s^T&=&{1\over 8\pi^{5/2}}\sum_{\lambda' = -1}^{+1}
\int_{-\infty}^{+\infty} dk^0 f_B(k^0)\theta[k_0^2-M^2+(2|\lambda'|-1)eB]
\left[\int_0^\infty \!{ds\over s^{5/2}}e^{-s (k_0^2-M^2)}\right.
\nonumber \\
&\times &
\left.
 e^{-2\lambda' eBs}{eBs\over \sinh (eBs)}
-{2\over \pi^{1/2}}\sum^\infty_{n=1}(-1)^n\left({eB\over n}\right)^{3/2}
\sin\left({\pi\over 4}-{n\pi\over eB}(k_0^2-M^2)\right)\right]
\label{LT2}
\end{eqnarray}
which can be equivalently written as
\begin{equation}
{\cal{L}}_W^T +\sum_s{\cal{L}}_s^T={eB\over 4\pi^{2}}
\sum_{n=0}^\infty\sum_{\lambda' = -1}^{+1}
\int_{-\infty}^\infty dk {k^2\over E_{n,\lambda'}}[f_B(E_{n,\lambda'})+f_B(-E_{n,\lambda'})]\,.
\label{LT5}
\end{equation}
where the Landau energy levels $E_{n,\lambda'}$ are given by Equation (\ref{Enl}).
The partition function $Z_W(B,T)$ and free energy
$F_W(B,T)$ of a $W$ boson plasma placed in a volume $V$ in a constant magnetic field $B$
is obtained immediately
\begin{equation}
{\ln Z_W(B,T)\over \beta V}={ F_W(B,T)\over V}={\cal{L}}_W^T +\sum_s{\cal{L}}_s^T\,.
\label{Zw}
\end{equation}

The lepton and quark fields contribution is 
\begin{eqnarray}
\sum_f {\cal{L}}_f^T &=&\sum_f{N_f\over 4\pi^{5/2}}
\int_{-\infty}^{+\infty} dk^0 f_F(k^0)\theta(k_0^2-m^2_f)\left[\int_0^\infty \!{ds\over s^{5/2}}
e^{-s (k_0^2-m^2_f)}eq_fBs \coth (eq_fBs)\right.
\nonumber \\
&-&
 \left. {2\over \pi^{1/2}}\sum^\infty_{n=1}\left({e|q_f|B\over n}\right)^{3/2}
\sin\left({\pi\over 4}-{n\pi\over e|q_f|B}(k_0^2-m^2_f)\right)\right]\,,
\label{LfT8}
\end{eqnarray}
where $m_f$ is the fermion mass. The summation index $f$ runs over the charged lepton 
fields $e,\mu, \tau$ and the quark fields
$u,d,c,s,t,b$ and $N_f=1$ and $q_f=1$ for each of the three leptons and, taking into account the quark
charges and colors, $N_f=3$ and $q_f=-2/3$ or $q_f=1/3$ for the six quarks. Equation (\ref{LfT8}) can also
be written as
\begin{equation}
\sum_f{\cal{L}}_f^T =\sum_f N_f{e|q_f| B\over 4\pi^{2}}
\sum_{n=0}^\infty\sum_{\lambda = 1}^2
\int_{-\infty}^\infty dk {k^2\over \omega_{n,\lambda}}[f_F(\omega_{n,\lambda})+f_F(-\omega_{n,\lambda})]\,,
\label{LfT9}
\end{equation}
where the Landau energy levels for a fermion are
\begin{equation}
\omega_{n,\lambda}=\sqrt{m^2_f+k^2+2e|q_f|B(n+\lambda-1)}\,.
\label{LfT10}
\end{equation}

The fermionic contribution to the vacuum part of ${\cal{L}}^{(1)}$ is \cite{heisenberg}
\begin{equation}
\sum_f{\cal{L}}_f^0 =-\sum_f {N_f\over 8\pi^2}
\int_0^\infty \!{ds\over s^3}\,\,
e^{-s m^2_f}\left[eq_fBs \coth (eq_fBs)-1-{(eq_fBs)^2\over 3}\right]\,,
\label{Lfv6}
\end{equation}
and the $W$ and scalar contribution is \cite{erdasfeld}
\begin{equation}
{\cal{L}}_W^0 +\sum_s{\cal{L}}_s^0 = {1\over 16\pi^2}
\int_0^\infty \!\!{ds\over s^3}\,\,
e^{-sM^2}\left[{eBs\over \sinh(eBs)} (1+e^{2eBs}+e^{-2eBs})-3-{7\over 2}(eBs)^2\right].
\label{LWv6}
\end{equation}
The renormalization procedure that was used to obtain $\sum_f{\cal{L}}_f^0$ and ${\cal{L}}_W^0 +\sum_s{\cal{L}}_s^0$
leads to the following value of the QED $\beta$ function for the electroweak theory
\begin{equation}
\beta^{EW}={2\alpha\over 3\pi}\sum_f {N_f}q^2_f-{7\alpha\over 2\pi}\,,
\label{beta}
\end{equation}
where the summation is the contribution to $\beta^{EW}$ from the fermions of the theory and the second term is the contribution
from the $W$ and the scalar fields.
%%%%%%%%%%%%%%%%%%%%%%%%%%%%%%%%%%%%%%%%%%%%%%%%%%%%%%
\section{Discussion and conclusions}
\label{4}
The effective electromagnetic lagrangian at finite temperature and density ${\cal{L}}^{(1)}$ 
obtained in the previous section contains several dimensionful parameters $T$, $B$, the 
charged fermion and gauge boson masses and the fermion chemical potentials.
Some of these parameters can be large or small compared to each other, and I 
will discuss some of these limits
that I find most interesting.

First I analyze the strong field limit and investigate the case of $eB\gg m^2_e$, which is the natural scale 
for the magnetic field strength required to significantly influence
quantum processes. Magnetic fields of this magnitude are present in many
astrophysical sites, and fields as large as
$B_e=m^2_e/e$ or larger can arise in magnetars, among others. In the literature 
\cite{elmfors1,elmfors2} the 
thermal and density corrections to the effective lagrangian have been neglected when compared to ${\cal{L}}_f^0$
for $f=e$ which, in the strong field limit and to leading order is
\begin{equation}
{\cal{L}}_e^0 \simeq {(eB)^2\over 24\pi^2}\ln \left({eB\over m^2_e}\right)\,.
\label{Lf0high1}
\end{equation}
However a more accurate calculation of ${\cal{L}}_f^0$ presented in the Appendix, where the next to leading order 
term is derived, shows that 
\begin{equation}
{\cal{L}}_e^0 ={(eB)^2\over 24\pi^2}\left[
\ln \left({eB\over 6 m^2_e}\right)-{1\over 2}\right]\,,
\label{Lf0high2}
\end{equation}
and therefore ${\cal{L}}_e^0\simeq 0$ when $B=10B_e$ and, for such values of $B$, the lightest quark provides
the main contribution to the vacuum part of the effective lagrangian ${\cal{L}}^0$, which can be written as
\begin{equation}
\sum_f{\cal{L}}_f^0+ {\cal{L}}_W^0 +\sum_s{\cal{L}}_s^0 \simeq {2(eB)^2\over 1,215\pi^2}
\left({eB\over m^2_u}\right)^2\,,
\label{Lf0high3}
\end{equation}
where $m_u$ is the $u$-quark mass. In this scenario, the thermal and density corrections to the effective lagrangian 
might be much larger than the vacuum part. For example, in the case of a $CP$-symmetric plasma with $eB\gg T^2$ 
and  $eB\gg m^2_e$, 
only the lowest Landau levels contributes to ${\cal{L}}_e^T$ and I use Equation (\ref{LfhighT}) to write
\begin{equation}
{\cal{L}}_e^T={eB\over 2\pi^2}\int_0^\infty {dk\over\omega_e}{k^2\over e^{\beta\omega_e}+1}\,,
\label{LfThigh1}
\end{equation}
where $\omega_e=\sqrt{k^2+m^2_e}$. An accurate analytical and numerical evaluation of the integral appearing in Equation
(\ref{LfThigh1}) yields
\begin{equation}
\int_0^\infty {dk\over\omega_e}{k^2\over e^{\beta\omega_e}+1}=\cases{{\pi^2\over 12}T^2+{m^2_e\over 4}
\left[\ln\left({m_e\over \pi T}\right)+\gamma_E-{1\over 2}\right]&
 for $T\ge m_e\,\,$;
 \cr \sqrt{\pi\over 2}\,T^2\left(\sqrt{m_e\over T}+{1\over 6}\right)e^{-m_e/T}& for $T\le m_e\,\,$,\cr}
\label{LfThigh2}
\end{equation}
where $\gamma_E$ is the Euler-Mascheroni constant. A numerical evaluation of ${\cal{L}}_e^T$ using Equations 
(\ref{LfThigh1}) and (\ref{LfThigh2}) shows that, for a $CP$-symmetric plasma with $B\sim 10 B_e$, the thermal 
correction given by Equation (\ref{LfThigh1}) is larger than the vacuum part of the effective lagrangian 
of Equation (\ref{Lf0high3}) for $T\ge m_e/6$ and, for $T = m_e$, I find that the thermal correction dominates
with ${\cal{L}}_e^T\sim 50{\cal{L}}^0$.

It is also interesting to investigate the case of a finite density medium where $\mu_e$ is the electron chemical potential.
For a finite density medium with strong magnetic field $eB\gg m^2_e\sim \mu^2_e-m^2_e \gg T^2$, the electron 
occupation number is $f_F(\omega_e)\simeq \theta(\omega_e)\theta((\omega_e-\mu_e)$ and I obtain from
Equation (\ref{LfhighT})
\begin{equation}
{\cal{L}}_e^T={eB\over 4\pi^2}\left[\mu_e\sqrt{\mu^2_e-m^2_e}-m^2_e\ln\left({\mu_e+\sqrt{\mu^2_e-m^2_e}\over m_e}\right)\right]\,.
\label{LfTmu1}
\end{equation}
It is neither intuitive nor convenient to express ${\cal{L}}_e^T$ in terms of the chemical potential,
it is better to use the difference between the electron and
positron number densities $n_e-n_{\bar e}$, which is related to $\mu_e$ by
\begin{equation}
n_e-n_{\bar e}={\partial{\cal{L}}_e^T\over \partial \mu_e}\,,
\label{LfTmu2}
\end{equation}
and therefore in a strong magnetic field it is given by
\begin{equation}
n_e-n_{\bar e}={eB\over 2\pi^2}\sqrt{\mu^2_e-m^2_e}\,.
\label{LfTmu3}
\end{equation}
Using the last relation, I eliminate $\mu$ from Equation (\ref{LfTmu1}) and rewrite it in the more transparent way
\begin{equation}
{\cal{L}}_e^T={ m_e\over 2}(n_e-n_{\bar e})\left[\sqrt{1+x^2}-{\ln\left(\sqrt{1+x^2}+x\right)\over x}\right]\,,
\label{LfTmu4}
\end{equation}
where 
\begin{equation}
x=2\pi^2{(n_e-n_{\bar e})\over eB m_e}
\label{x}
\end{equation}
is a dimensionless parameter. I find that in a medium with magnetic field $B\sim 10B_e$, temperature $T\ll m_e$ and 
density $\rho\sim 10^{10} {\rm Kg/m^3}$ the finite density correction to the effective lagrangian 
obtained from Equation (\ref{LfTmu4}), dominates 
over the vacuum part ${\cal{L}}_e^T\gg{\cal{L}}^0$. These conditions are believed to be common
in neutron stars and magnetars.

Next I investigate ${\cal{L}}_W^T +\sum_s{\cal{L}}_s^T$, which is the 
thermal correction to the effective lagrangian from the $W$ and the scalars, and 
compare it to the vacuum part ${\cal{L}}_W^0 +\sum_s{\cal{L}}_s^0$, in the case of magnetic field
$B\ll M^2/e$, where $M$ is the $W$ mass. In this limit, Equation (\ref{LT2}) becomes
\begin{equation}
{\cal{L}}_W^T +\sum_s{\cal{L}}_s^T={1\over \pi^2}\int^\infty_0{dk\over \omega_W}
{k^4\over e^{\beta \omega_W}-1}
+{7\over 8\pi^2}(eB)^2\int^\infty_0{dk\over \omega_W}
{1\over e^{\beta \omega_W}-1}\,,
\label{Wft1}
\end{equation}
where $\omega_W=\sqrt{k^2+M^2}$. Notice that the first term, independent of $B$, 
is the field independent thermal 
correction to the lagrangian and that the term with the sum over $n$ of Equation (\ref{LT2})
is neglected because, for $eB\ll M^2$, this term is proportional to $(eB)^3/M^2$. 
After an accurate evaluation of the integrals appearing in Equation
(\ref{Wft1}) I obtain
\begin{equation}
\int_0^\infty {dk\over\omega_W}{k^4\over e^{\beta\omega_W}-1}=\cases{{\pi^4 T^4\over 15}
-{\pi^2 T^2M^2\over 4}+{\pi T M^3\over 2}&
 for $T\ge {M\over \sqrt{2}}\,\,$;
 \cr 3\sqrt{\pi\over 2}T^{5/2}M^{3/2}\left[1+\left({T\over M}\right)^{1/4}+
 {2\over 3}\left({T\over M}\right)^{1/2}\right]
 e^{-M/T}& for $T\le {M\over \sqrt{2}}\,\,$,\cr}
\label{Wft2a}
\end{equation}
and 
\begin{equation}
\int_0^\infty {dk\over\omega_W}{1\over e^{\beta\omega_W}-1}=\cases{{\pi T\over 2M}
+{1\over 2}
\left[\ln\left({M\over 4\pi T}\right)+\gamma_E-{1\over 2}\right]&
 for $T\ge \sqrt{10}M\,\,$;
 \cr \sqrt{\pi T\over 2M}\left(1+{1\over 4}\sqrt{T\over M}\right)
 e^{-M/T}& for $T\le \sqrt{10}M\,\,$.\cr}
\label{Wft2}
\end{equation}
For $eB\ll M^2$ the $W$ and scalar contribution to the vacuum part of the effective lagrangian is
obtained easily from Equation (\ref{LWv6})
\begin{equation}
{\cal{L}}_W^0 +\sum_s{\cal{L}}_s^0 = {29\over 640 \pi^2}{(eB)^4\over M^4}\,,
\label{Wft3}
\end{equation}
and we see that, for 
\begin{equation}
T\sim {M\over 2\ln(M^2/eB)}
\end{equation}
or higher, the $B$-dependent term of Equation (\ref{Wft1}) dominates over $
{\cal{L}}_W^0 +\sum_s{\cal{L}}_s^0$, as given by Equation (\ref{Wft3}). 
On the other hand, the $W$ and scalars 
contribution becomes a significant part of the lagrangian only for $T\sim M$ and $m^2_e\ll
eB\ll M^2$.

Last I will investigate the implications of my results on the effective QED coupling in the
medium. Due to the scale invariance of $eB$ we can define an effective coupling constant from
${\cal L}^{(1)}$ as \cite{schwinger,elmfors2,chodos}
\begin{equation}
-{1\over e^2(B,T,\mu)}={1\over eB}{\partial {\cal L}^{(1)}\over \partial(eB)}\,,
\label{eff1}
\end{equation}
from which we obtain the effective electromagnetic fine structure constant $\alpha(B,T,\mu)
 =e^2(B,T,\mu)/4\pi$
\begin{equation}
{1\over \alpha(B,T,\mu)}={1\over \alpha}-{1\over \alpha B}{\partial ({\cal L}^{0}+{\cal L}^{T})
\over \partial B}\,.
\label{eff2}
\end{equation}
In the limit when
$eB=0$, I find that the effective coupling $\alpha(T,\mu)=\alpha(B=0,T,\mu)$ is given by
\begin{equation}
{1\over \alpha(T,\mu)}={1\over \alpha}-{2\over 3\pi}\sum_f N_fq^2_f
\int_0^\infty {dk\over\omega_f}[f_F^+(\omega_f)+f_F^-(-\omega_f)]-
{7\over 2\pi}\int_0^\infty {dk\over\omega_W}{1\over e^{\beta \omega_W}-1}\,,
\label{eff3}
\end{equation}
with $\omega_f=\sqrt{k^2+m^2_f}$. When $T=0$, I find an effective coupling  $\alpha(\mu)=\alpha(T=0,\mu)$
that is
\begin{equation}
{1\over \alpha(\mu)}={1\over \alpha}-{2\over 3\pi}\sum_f N_fq^2_f
\ln\left({|\mu_f|\over m_f}+\sqrt{{\mu_f^2\over m^2_f}-1}\right)\,,
\label{eff4}
\end{equation}
where $\mu_f$ is the chemical potential of the fermion $f$. Clearly, in an electrically neutral medium at low temperature, the
summation will be over $e$, $u$, and $d$ only, since only their chemical potential will be relevant. 

In a CP-symmetric medium where the 
temperature is very high but still below the critical temperature of the Weinberg-Salam model, $M\ge T\gg m_b$ where $m_b$ is the
mass of the $b$ quark,
I find the following behavior of the corresponding effective
coupling $\alpha(T)=\alpha(T,\mu~=~0)$
\begin{equation}
{1\over \alpha(T)}={1\over \alpha}-{2\over 3\pi}\sum_{f'} N_{f'}q^2_{f'}
\ln\left({T\over m_{f'}}\right)-
{16\over 9\pi}\int_0^\infty {dk\over\omega_t}{1\over e^{\beta \omega_t}+1}-
{7\over 2\pi}\int_0^\infty {dk\over\omega_W}{1\over e^{\beta \omega_W}-1}\,,
\label{eff5}
\end{equation}
where the sum is over all fermions except for the quark $t$ and $\omega_t=\sqrt{k^2+m^2_t}$. The value of the
last integral appearing above
is given by Equation (\ref{Wft2}) and, after an accurate analytical and numerical 
evaluation, I find the value of the other integral 
\begin{equation}
\int_0^\infty {dk\over\omega_t}{1\over e^{\beta \omega_t}+1}=\cases{
-{1\over 2}
\left[\ln\left({m_t\over \pi T}\right)+\gamma_E\right]&
 for $T\ge m_t\,\,$;
 \cr \sqrt{\pi T\over 2m_t}\left(1-{1\over 4}\sqrt{T\over m_t}\right)
 e^{-m_t/T}& for $T\le m_t\,\,$.\cr}
\label{eff6}
\end{equation}

Finally I investigate the behavior of $\alpha(B)=\alpha(T=0,\mu=0, B)$ when $eB\rightarrow M^2$ and $T=0$, $\mu=0$.
Under these conditions, I obtain the following form of the $W$ and scalar contribution to the vacuum effective lagrangian
\begin{equation}
{\cal{L}}_W^0 +\sum_s{\cal{L}}_s^0 \simeq {1\over 16 \pi^2}\left[\left({7\over 2}-
{7\over 2}\ln 2-12\zeta'(-1)\right)e^2B^2+2eB(M^2-eB)\ln\left({M^2-eB\over M^2+eB}\right)
\right]
\label{eff7}
\end{equation}
where a constant term proportional to $M^4$ has been discarded and $\zeta'(-1)=-0.16541$ is the derivative of 
the Riemann zeta function. Notice the logarithmic branch point in the last term 
at $eB=M^2$, which indicates that the effective lagrangian picks up an 
imaginary part when $eB>M^2$ and confirms the well-known result \cite{ambjorn1, ambjorn2, nielsen}
concerning the instability of the vacuum.
Using Equation (\ref{eff7}), I find the following behavior of $\alpha(B)$ when $eB\rightarrow M^2$
\begin{equation}
{1\over \alpha(B)}={1\over \alpha}-{1\over 3\pi}\sum_{f'} N_{f'}q^2_{f'}
\ln\left({eB\over m_{f'}^2}\right)-
{1\over 2\pi}\ln\left({M^2+eB\over M^2-eB}\right)\,,
\label{efflast}
\end{equation}
where the summation is over all fermions with the exception of the quark $t$, whose 
contribution is negligible since it is heavier than the $W$. The contribution of the $W$ and scalars to the
effective coupling constant of Equation (\ref{efflast}) is the last term, with the logarithmic branch point 
at $eB=M^2$.
%%%%%%%%%%%%%%%%%%%%%%%%%%%%%%%%%%%%%%%%%%%%%%%%%%%%%%%%%%%%%%%%%%%%%
\appendix*
\section{}

In this appendix I evaluate the Heisenberg-Euler vacuum effective lagrangian and the Weisskopf vacuum effective lagrangian 
of scalar QED in the case of strong magnetic field $eB\gg m^2$. First I evaluate ${\cal L}^0_f$, 
the vacuum effective lagrangian for spinor QED of Equation (\ref{Lf03}). I introduce a 
regulator to evaluate the integrals, change the variable of integration from $s$ to $z=eBs$ and write
\begin{equation}
{\cal{L}}_f^0 =-\lim_{\epsilon \rightarrow 0} {(eB)^{2-\epsilon}\over 8\pi^2}[I_1(\epsilon,x)-I_2(\epsilon,x)]\,,
\label{app1}
\end{equation}
where 
\begin{equation}
I_1(\epsilon,x)=\int_0^\infty \!{dz\over z^{3-\epsilon}}\,\,
e^{-z x}\left[z \coth z-1\right]\,,
\label{app2}
\end{equation}
 
\begin{equation}
I_2(\epsilon,x)={1\over 3}\int_0^\infty \!{dz\over z^{1-\epsilon}}\,\,
e^{-z x}\,,
\label{app3}
\end{equation}
and $x=m^2/eB$. When $x\ll 1$, I can take $e^{-zx}\simeq 1$ inside the integral of Equation (\ref{app2}),
use the following series expansion of the hyperbolic cotangent
\begin{equation}
\coth z={1\over z} + 2z\sum_{n=1}^\infty {1\over (n\pi)^2+z^2}\,,
\end{equation}
and find
\begin{equation}
I_1(\epsilon,x)=2\sum_{n=1}^\infty\int_0^\infty \!{dz\over z^{1-\epsilon}}\,\,
 {1\over (n\pi)^2+z^2}\,.
\label{app4}
\end{equation}
I evaluate this integral and obtain
\begin{equation}
I_1(\epsilon,x)={\pi^{\epsilon-2}}\zeta (2-\epsilon) B\left({\epsilon\over 2},1-
{\epsilon\over 2}\right)\,,
\label{app5}
\end{equation}
where $\zeta (2-\epsilon)$ is the Riemann zeta function and 
$B\left({\epsilon\over 2},1-{\epsilon\over 2}\right)$ is the Euler beta function. I evaluate $I_2(\epsilon,x)$ exactly
and find
\begin{equation}
I_2(\epsilon,x)={1\over 3}x^{-\epsilon}\Gamma(\epsilon)\,,
\label{app6}
\end{equation}
where $\Gamma(\epsilon)$ is the Euler gamma function. I insert into Equation (\ref{app1}) the 
values of $I_1$ and $I_2$ that I have found, take the limit $\epsilon\rightarrow 0$ and obtain
\begin{equation}
{\cal{L}}_f^0 =-{(eB)^2\over 8\pi^2}\left[
{1\over 3} \ln (\pi x)+{1\over 3}\gamma_E-{2\over \pi^2}\zeta'(2)\right]\,,
\label{app7}
\end{equation}
where $\gamma_E=0.5772$ and $\zeta'(2)=-0.9375$ is the first derivative of the
Riemann zeta function. Finally, I can significantly simplify Equation (\ref{app7})
using the interesting numerical fact \cite{erdas} that 
${6\over \pi^2}\zeta'(2)-\log \pi-\gamma_E-=-2.2918=-\ln 6 -{1\over 2} $, and write
\begin{equation}
{\cal{L}}_f^0 ={(eB)^2\over 24\pi^2}\left[
\ln \left({eB\over 6 m^2}\right)-{1\over 2}\right]\,.
\label{app8}
\end{equation}
I proceed in a similar way when calculating the Weisskopf effective lagrangian 
of scalar QED for $eB\gg m^2$, where $m$ is the mass of the scalar field. While this result will not be used 
in this paper, I believe it is of some interest, considering the attention received by the Weisskopf 
effective lagrangian throughout the years. I start with Equation (\ref{Ls01}) for ${\cal{L}}_s^0$ and,
after introducing the regulator and changing the 
integration variable to $z$, I write
\begin{equation}
{\cal{L}}_s^0 =\lim_{\epsilon \rightarrow 0} {(eB)^{2-\epsilon}\over 16\pi^2}[I_3(\epsilon,x)+{1\over 2}I_2(\epsilon,x)]\,,
\label{app9}
\end{equation}
where $I_2(\epsilon,x)$ is given in Equation (\ref{app3}) and 
\begin{equation}
I_3(\epsilon,x)=\int_0^\infty \!{dz\over z^{3-\epsilon}}\,\,
e^{-z x}\left[{z\over \sinh z}-1\right]\,.
\label{app10}
\end{equation}
After setting $e^{-zx}\simeq 1$ and
using the following series expansion of the hyperbolic cosecant
\begin{equation}
{1\over\sinh z}={1\over z} + 2z\sum_{n=1}^\infty {(-1)^n\over (n\pi)^2+z^2}\,,
\end{equation}
I evaluate $I_3(\epsilon,x)$ and obtain
\begin{equation}
I_3(\epsilon,x)={\pi^{\epsilon-2}}(2^{\epsilon -1}-1)\zeta (2-\epsilon) B\left({\epsilon\over 2},1-
{\epsilon\over 2}\right)\,.
\label{app11}
\end{equation}
After inserting this value of $I_3(\epsilon,x)$ and the value of $I_2(\epsilon,x)$ obtained previously 
inside Equation (\ref{app9}), I take $\epsilon \rightarrow 0$ and find
\begin{equation}
{\cal{L}}_s^0 ={(eB)^2\over 16\pi^2}
\left[ -{1\over 6} \ln \left({\pi\over 2} x\right)
-{1\over 6}\gamma_E-{1\over \pi^2}\zeta'(2)
\right]\,,
\label{app12}
\end{equation}
which, using the aforementioned interesting numerical fact, I rewrite as
\begin{equation}
{\cal{L}}_s^0 ={(eB)^2\over 96\pi^2}\left[
\ln \left({eB\over 3 m^2}\right)-{1\over 2}\right]\,.
\label{app13}
\end{equation}

%%%%%%%%%%%%%%%%%%%%%%%%%%%%%%%%%%%%%%%%%%%%%%%%%%%%%%%%%%%%%%%%%%%%%

%%%%%%%%%%%%%%%%%%%%%%%%%%%%%%%%%%%%%%%%%%%%%%%%%%%%%%%%%%%%%%%%%%%%%

\begin{thebibliography}{99}

\bibitem{Duncan}
  R.~C.~Duncan and C.~Thompson,
  Astrophys.\ J.\  {\bf 392}, L9 (1992).

\bibitem{Thompson1}
  C.~Thompson and R.~C.~Duncan,
  Mon.\ Not.\ Roy.\ Astron.\ Soc.\  {\bf 275}, 255 (1995).

\bibitem{Thompson2}
  C.~Thompson and R.~C.~Duncan,
  Astrophys.\ J.\  {\bf 473}, 322 (1996).
  
\bibitem{Brandenburg}
  A.~Brandenburg, K.~Enqvist and P.~Olesen,
  Phys.\ Rev.\  D {\bf 54}, 1291 (1996).

\bibitem{Joyce}
  M.~Joyce and M.~E.~Shaposhnikov,
  Phys.\ Rev.\ Lett.\  {\bf 79}, 1193 (1997).

\bibitem{grasso}
  D.~Grasso and H.~R.~Rubinstein,
  Phys.\ Rept.\  {\bf 348}, 163 (2001).

  \bibitem{heisenberg}
  W.~Heisenberg and H.~Euler,
  %``Consequences of Dirac's theory of positrons,''
  Z.\ Phys.\  {\bf 98}, 714 (1936)

\bibitem{Weisskopf:1996bu}
  V.~Weisskopf,
  %``The Electrodynamics Of The Vacuum Based On The Quantum Theory Of The
  %Electron,''
%\href{http://www.slac.stanford.edu/spires/find/hep/www?irn=3773230}{SPIRES entry}
{\it  In *Miller, A.I.: Early quantum electrodynamics* 206-226}

 \bibitem{karplus}
  R.~Karplus and M.~Neuman,
  %``The scattering of light by light,''
  Phys.\ Rev.\  {\bf 83}, 776 (1951).

\bibitem{schwinger}
  J.~S.~Schwinger,
  %``On gauge invariance and vacuum polarization,''
  Phys.\ Rev.\  {\bf 82}, 664 (1951).

\bibitem{Dunne:2008kc}
  G.~V.~Dunne,
  %``New Strong-Field QED Effects at ELI: Nonperturbative Vacuum Pair
  %Production,''
  Eur.\ Phys.\ J.\  D {\bf 55}, 327 (2009)

\bibitem{Hebenstreit:2009km}
  F.~Hebenstreit, R.~Alkofer, G.~V.~Dunne and H.~Gies,
  %``Momentum signatures for Schwinger pair production in short laser pulses
  %with sub-cycle structure,''
  Phys.\ Rev.\ Lett.\  {\bf 102}, 150404 (2009)
  
\bibitem{dittrich2}
  W.~Dittrich and H.~Gies,
  %``Probing the quantum vacuum. Perturbative effective action approach in
  %quantum electrodynamics and its application,''
  Springer Tracts Mod.\ Phys.\  {\bf 166}, 1 (2000).

 \bibitem{elmfors1}
  P.~Elmfors, D.~Persson and B.~S.~Skagerstam,
  %``QED effective action at finite temperature and density,''
  Phys.\ Rev.\ Lett.\  {\bf 71}, 480 (1993)

\bibitem{elmfors2}
  P.~Elmfors, D.~Persson and B.~S.~Skagerstam,
  %``Real time thermal propagators and the QED effective action for an external
  %magnetic field,''
  Astropart.\ Phys.\  {\bf 2}, 299 (1994)

\bibitem{elmfors3}
  P.~Elmfors, P.~Liljenberg, D.~Persson and B.~S.~Skagerstam,
  %``Thermal versus vacuum magnetization in QED,''
  Phys.\ Rev.\  D {\bf 51}, 5885 (1995)

\bibitem{chodos}
  A.~Chodos, D.~A.~Owen and C.~M.~Sommerfield,
  %``STRONG FIELD DEPENDENCE OF THE FINE STRUCTURE CONSTANT,''
  Phys.\ Lett.\  B {\bf 212}, 491 (1988).

\bibitem{Kim:2008em}
  S.~P.~Kim, H.~K.~Lee and Y.~Yoon,
  %``Schwinger Pair Production at Finite Temperature in QED,''
  Phys.\ Rev.\  D {\bf 79}, 045024 (2009)

\bibitem{Kim:2010qq}
  S.~P.~Kim, H.~K.~Lee and Y.~Yoon,
  %``Nonperturbative QED Effective Action at Finite Temperature,''
  arXiv:1006.0774 [hep-th].

\bibitem{Das:2009ci}
  A.~Das and J.~Frenkel,
  %``Effective actions at finite temperature,''
  Phys.\ Rev.\  D {\bf 80}, 125039 (2009)

\bibitem{Das:2009vs}
  A.~Das and J.~Frenkel,
  %``Finite Temperature Effective Actions,''
  Phys.\ Lett.\  B {\bf 680}, 195 (2009)

\bibitem{Teo:2008ah}
  L.~P.~Teo,
  %``Finite temperature Casimir effect in spacetime with extra compactified
  %dimensions,''
  Phys.\ Lett.\  B {\bf 672}, 190 (2009)

\bibitem{Lim:2009zza}
  S.~C.~Lim and L.~P.~Teo,
  %``Repulsive Casimir force for electromagnetic fields with mixed boundary
  %conditions,''
  Int.\ J.\ Mod.\ Phys.\  A {\bf 24}, 3455 (2009).

\bibitem{Milton:2010yw}
  K.~A.~Milton, J.~Wagner, P.~Parashar and I.~Brevik,
  %``Casimir energy, dispersion, and the Lifshitz formula,''
  Phys.\ Rev.\  D {\bf 81}, 065007 (2010)

\bibitem{Frank:2007jb}
  M.~Frank, I.~Turan and L.~Ziegler,
  %``The Casimir force in Randall Sundrum models,''
  Phys.\ Rev.\  D {\bf 76}, 015008 (2007)

\bibitem{Rypestol:2009pe}
  M.~Rypestol and I.~Brevik,
  %``Finite Temperature Casimir Effect in Randall-Sundrum Models,''
  New J.\ Phys.\  {\bf 12}, 013022 (2010)

\bibitem{Cannellos:2002ex}
  J.~Cannellos, E.~J.~Ferrer and V.~de la Incera,
  %``Hypermagnetic field effects in the thermal bath of chiral fermions,''
  Phys.\ Lett.\  B {\bf 542}, 123 (2002)

\bibitem{Piccinelli:2004eu}
  G.~Piccinelli and A.~Ayala,
  %``Electroweak baryogenesis and primordial hypermagnetic fields,''
  Lect.\ Notes Phys.\  {\bf 646}, 293 (2004)
    
\bibitem{Sanchez:2006tt}
  A.~Sanchez, A.~Ayala and G.~Piccinelli,
  %``Effective potential at finite temperature in a constant hypermagnetic
  %field: Ring diagrams in the standard model,''
  Phys.\ Rev.\  D {\bf 75}, 043004 (2007)
  
\bibitem{Ambjorn:1992ca}
  J.~Ambjorn and P.~Olesen,
  %``Electroweak magnetism, W condensation and antiscreening,''
  arXiv:hep-ph/9304220.
  %%CITATION = HEP-PH/9304220;%%

\bibitem{dittrich}
  W.~Dittrich and M.~Reuter,
  %``Effective Lagrangians In Quantum Electrodynamics,''
  Lect.\ Notes Phys.\  {\bf 220}, 1 (1985).

\bibitem{erdasfeld}
  A.~Erdas and G.~Feldman,
  %``MAGNETIC FIELD EFFECTS ON LAGRANGIANS AND NEUTRINO SELFENERGIES IN THE
  %SALAM-WEINBERG THEORY IN ARBITRARY GAUGES,''
  Nucl.\ Phys.\  B {\bf 343}, 597 (1990).

\bibitem{ambjorn1}
  J.~Ambjorn and P.~Olesen,
  %``ON ELECTROWEAK MAGNETISM,''
  Nucl.\ Phys.\  B {\bf 315}, 606 (1989)
 
\bibitem{ambjorn2}
  J.~Ambjorn and P.~Olesen,
  %``Antiscreening Of Large Magnetic Fields By Vector Bosons,''
  Phys.\ Lett.\  B {\bf 214}, 565 (1988)

\bibitem{nielsen}
  N.~K.~Nielsen and P.~Olesen,
  %``An Unstable Yang-Mills Field Mode,''
  Nucl.\ Phys.\  B {\bf 144}, 376 (1978).
 
 \bibitem{erdas}
  A.~Erdas,
  %``Neutrino self-energy in external magnetic field,''
  Phys.\ Rev.\  D {\bf 80}, 113004 (2009)

\end{thebibliography}
\end{document}